\documentclass[aps,pra,twocolumn,groupedaddress,showpacs]{revtex4-1}
\usepackage{graphicx}
\usepackage{longtable} 

\begin{document}


\title{Development of a compact HAPG crystal Von Hamos X-ray spectrometer for extended and diffused sources}


\author{A. Scordo}
\email[]{alessandro.scordo@lnf.infn.it}
\affiliation{Laboratori Nazionali di Frascati INFN, Frascati (Rome), Italy}

\author{C. Curceanu}
\affiliation{Laboratori Nazionali di Frascati INFN, Frascati (Rome), Italy}
\author{M. Miliucci}
\affiliation{Laboratori Nazionali di Frascati INFN, Frascati (Rome), Italy}
\author{F. Sirghi}
\affiliation{Laboratori Nazionali di Frascati INFN, Frascati (Rome), Italy}
\affiliation{Horia Hulubei National Institute of Physics and Nuclear Engineering (IFIN-HH), Magurele, Romania}
\author{J. Zmeskal}
\affiliation{Stefan-Meyer-Institut f\" ur subatomare Physik,Vienna, Austria}

\date{\today}

\begin{abstract}
Bragg spectroscopy is one of the best established experimental methods for high energy resolution X-ray measurements; however, this technique is limited to the measurement of photons produced from well collimated (tens of microns) or point-like sources and becomes quite inefficient for photons coming from extended and diffused sources. The possibility to perform simultaneous measurements of several energies is strongly demanded when low rate signals are expected and single angular scans require long exposure times.
A prototype of a high resolution and high precision X-ray spectrometer working also with extended isotropic sources, has been developed by the VOXES collaboration at INFN Laboratories of Frascati, using Highly Annealed Pyrolitic Graphite (HAPG) crystals in a ``semi''- Von Hamos configuration, in which the position detector is rotated with respect to the standard Von Hamos one, to increase the dynamic energy range. The aim is to deliver a cost effective system having an energy resolution at the level of eV for X-ray energies from about 2 keV up to tens of keV, able to perform sub-eV precision measurements with non point-like sources. 
The proposed spectrometer has possible applications in several fields, going from fundamental physics to quantum mechanics tests, synchrotron radiation and X-FEL applications, astronomy, medicine and industry. 
In particular, this technique is fundamental for a series of nuclear physics measurements like, for example, the energies of the exotic atoms radiative transitions which allow to extract fundamental parameters in the low energy QCD in the strangeness sector.
In this work, the working principle of the spectrometer is presented, together with the tests and the results, in terms of resolution and source, size obtained for $Fe(K_{alpha1,2})$, $Cu(K_{\alpha1,2})$, $Ni(K_{\beta})$, $Zn(K_{\alpha1,2})$, $Mo(K_{\alpha1,2})$ and $Nb(K_{\beta})$ lines.
\end{abstract}


\maketitle

\section{Introduction \label{intro}}

\noindent The possibility to perform high precision measurements of soft X-rays, strongly demanded in many fields of fundamental science, from particle and nuclear
physics to quantum mechanics, as well as in astronomy and in several applications using synchrotron light sources or X-FEL beams, in biology, medicine and industry, is still today a big challenge. 
These measurements are even more difficult when they have to be performed in accelerator environments where, depending on the different machines, various kinds of hadronic and electromagnetic backgrounds are present.
Typical large area spectroscopic detectors used for wide and isotropic targets in accelerator environments are solid state devices, like the Silicon Drift Detectors (SDDs), recently employed by the SIDDHARTA experiment \cite{Bazzi:2011zj} for exotic atoms transition lines measurements at the $DA\Phi NE$ $e^+e^-$ collider of the INFN National Laboratories of Frascati \cite{Gallo:2006yn}.  
The intrinsic resolution of such kind of detectors is nevertheless limited to $\simeq\,120\,eV$ FWHM by the electronic noise and the Fano Factor, making them unsuitable for those cases in which the photon energy has to be measured with a precision of below 1 eV.

\noindent Few eV resolutions have become achievable using superconducting microcalorimeters, like the Transition Edge Sensors developed at NIST \cite{Doriese:2017yn}, able to obtain few eV FHWM at 6 keV; 
in spite of this excellent resolution, these kind of detectors still have some limitations: a very small active area, prohibitively high costs of the complex cryogenic system needed 
to reach the operational temperature of $\simeq\,50\,mK$, and a response function which is still not properly under control.

\noindent As a third possibility, Bragg spectroscopy is one of the best established high resolution X-ray measurement techniques; however, when the photons emitted from extended sources (like a gaseous or liquid target) have to be measured, this method has been until now ruled out by the constraint to reduce the dimension of the target to a few tens of microns \cite{Legall2006}\cite{Barnsley2003}. 
Experiments performed in the past at the Paul Scherrer Institute (PSI), measuring pionic atoms \cite{Anagnostopoulos:2001nfa}\cite{Trassinelli:2016kki}, 
pioneered the possibility to combine Charged Coupled Device detectors (CCDs) with silicon crystals, but the energy range achievable with that system was limited to few keV 
due to the crystal structure, and the silicon low intrinsic reflection efficiency required the construction of a very large spectrometer.
The possibility to perform other fundamental measurements, like the precision determination of the $K^-$ mass measuring the radiative kaonic nitrogen transitions at the $DA\Phi NE$ collider \cite{Beer:2002ug}, has been also investigated, but the estimated efficiency of the proposed spectrometer was not sufficient to reach the required precision. 

\noindent Recently, the development of the Pyrolitic Graphite mosaic crystals \cite{Sanchez:1998}, renewed the interest on Bragg spectrometers as possible candidates also for millimetric isotropic sources X-ray measurements in accelerator environments.
Mosaic crystals consist in a large number of nearly perfect small pyrolitic graphite crystallites, randomly misoriented around the lattice main direction;
the FWHM of this random angular distribution is called mosaicity ($\omega_{FWHM}$) and it makes possible that even a photon not reaching the crystal with the exact Bragg energy-angle relation, 
can find a properly oriented crystallite and be reflected \cite{Gerlach:2015}. This, together with a lattice spacing constant of $3,514$ {\AA}, 
enables them to be highly efficient in diffraction in the 2-20 keV energy range, for the n=1 reflection order, while higher energies can be reached at higher reflection orders. 

\noindent Thanks to their production mechanism,  Higly Annealed Pyrolitic Graphite crystals (HAPG) can be realised with different ad-hoc geometries, making them suitable
to be used in the Von Hamos configuration \cite{VonHamos:1933}, combining the dispersion of a flat crystal with the focusing properties of cilindrically
bent crystals. 

\noindent The main problem to overcome is represented by the source size; Von Hamos spectrometers have been extensively used in the past providing very promising results in terms of spectral resolution \cite{Legall2006}\cite{Shevelko:2002}\cite{Zastrau:2013}\cite{Anklamm:2014}
but all the available works in literature report measurements done in conditions to have an effective source dimensions of some tens of microns; this configuration is achieved either with microfocused X-ray tubes, 
or with a set of slits and collimators placed before the target to minimize the activated area. 

\noindent In this work we investigate the possibility to combine HAPG crystals properties with the vertical focus of the Von Hamos configuration, to realize a spectrometer able to maintain a resolution in the order of $0,1\%$ (FWHM/E), for energies below $10\, keV$, and of $0,5\%$ up to $20\, keV$, using a source size ranging from $500\,\mu m$ to $2\, mm$ in the Bragg dispersion plane. In section \ref{setup} the experimental setup and the spectrometer geometry are presented, while in section \ref{measure} the experimental results obtained for the $Fe(K_{\alpha1,2})$, $Cu(K_{\alpha1,2})$, $Ni(K_{\beta})$, $Zn(K_{\alpha1,2})$, $Mo(K_{\alpha1,2})$ and $Nb(K_{\beta})$ line measurements are reported. 

\section{Spectrometer setup and geometry \label{setup}}
\subsection{Von Hamos geometry}

\begin{figure}[htbp]
\centering
\includegraphics[width=8.7cm]{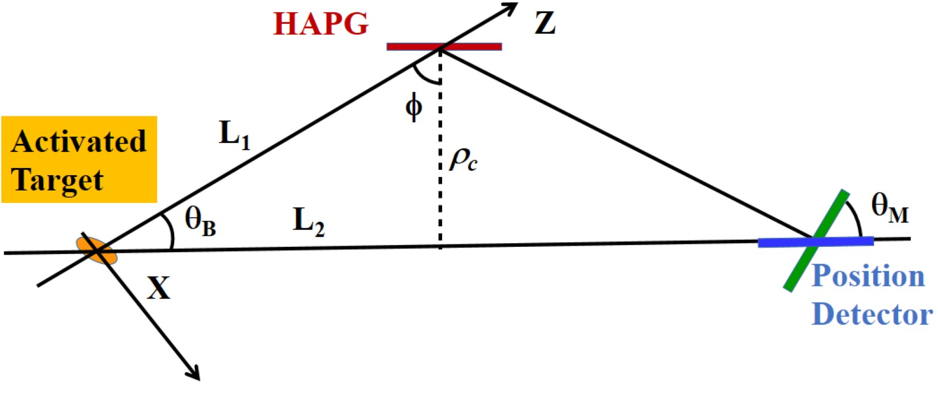}%
\caption{Von Hamos schematic geometry schematic of the dispersive plane (not in scale): the X-ray source is pictured in orange, the HAPG crystal in red and the position detector in blue and green for the standard and the ``semi'' Von Hamos configuration (see section \ref{semi}), respectively. $\rho_c$ is the crystal curvature radius, $\theta_B$ is the Bragg angle, $\phi=\pi-\theta_B$, $\theta_M$ is the position detector rotation angle with respect to standard VH configuration, $L_1$ is the source-crystal distance and $L_2$ is half the resulting source-detector distance. \label{vhl1l2}}
\end{figure}

\noindent The spectrometer configuration used in the measurements presented in this work is the Von Hamos one, in which the X-ray source and the position detector are placed on the axis of a cylindrically bent crystal (see fig. \ref{vhl1l2}; this geometrical scheme allows an improvement in the reflection efficiency due to the vertical focusing. 
As a consequence, for each X-ray energy the source-crystal ($L_1$) and the source-detector ($L_2$) distances are determined by the Bragg angle ($\theta_B$) and the curvature radius of the crystal ($\rho_c$): 

\begin{equation}
L_1 = \frac{\rho_c}{sin\theta_B}
\end{equation}

\begin{equation}
L_2 = L_1{sin\phi}
\end{equation}

\noindent In Fig. \ref{vhl1l2}, a schematic of the dispersive plane is shown where the X-ray source is sketched in orange, the HAPG crystal in red and the position detector in blue and green for the standard and the ``semi'' Von Hamos configuration (see section \ref{semi}), respectively. In the figure, $\rho_c$ is the crystal curvature radius, $\theta_B$ is the Bragg angle, $\phi=\pi-\theta_B$, $L_1$ is the source-crystal distance and $L_2$ is half of the resulting source-detector distance. 

\subsubsection{Spectrometer components}

\begin{figure}[htbp]
\centering
\includegraphics[width=8.7cm]{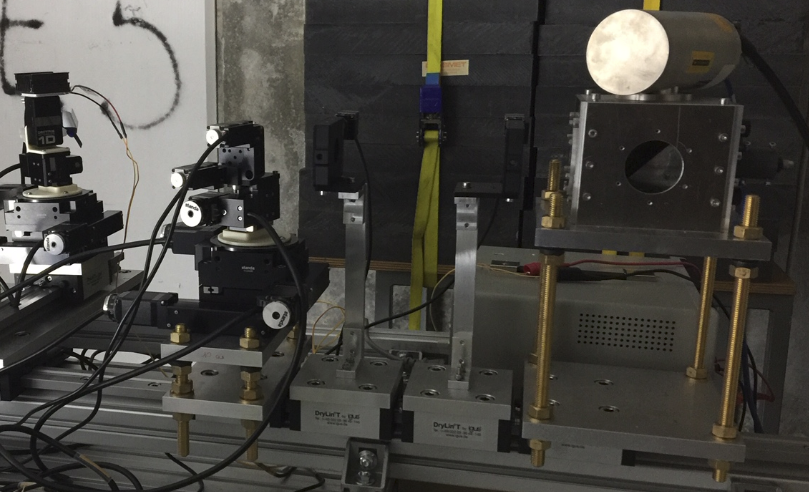}%
\caption{Spectrometer setup picture. \label{setpic}}
\end{figure}

\noindent The setup used to perform the reported measurements is shown in Fig. \ref{setpic}.

\noindent A single or multi element thin foil (target) is placed inside an aluminum box and activated by a XTF-5011 Tungsten anode X-ray tube, produced by OXFORD INSTRUMENTS, 
placed on top of the box; the center of the foil, placed on a $45^{\circ}$ rotated support prism, represents the origin of the reference frame in which Z is the direction of the characteristic photons emitted by 
the target and forming, with the X axis, the Bragg reflection plane, while Y is the vertical direction, along which primary photons generated by the tube are shot.
Two adjustable motorized slits (STANDA 10AOS10-1) with $1\,\mu m$ resolution are placed after the $5,9\,mm$ diameter circular exit window of the aluminum box in order to
shape the outcoming X-ray beam. The various used HAPG crystals, having a thickness of $100\,\mu m$ and a mosaicity of $(0,1\pm0,03)^{\circ}$, are deposited on different curvature 
radii Thorlabs N-BK7 $30\times\,32\,mm^2$ uncoated Plano-Concave Cylindrical lenses, 
held by a motorized mirror mount (STANDA 8MUP21-2) with a double axis $\ll\,1\,arcsec$ resolution and coupled
to a $0,01\,\mu m$ motorized vertical translation stage (STANDA 8MVT40-13), a $4,5\,arcsec$ resolution rotation stage (STANDA 8MR191-28) and two motorized $0,156\,\mu m$ linear translation stages (STANDA 8MT167-25LS).
The position detector is a commercial MYTHEN2-1D 640 channels strip detector produced by DECTRIS (Zurich, Switzerland), having an active area of $32\times8\,mm^2$ and 
whose strip width and thickness are, respectively, $50\,\mu m$ and $420\,\mu m$; the MYTHEN2-1D detector is also coupled to a positioning motorized system identical to the one for the HAPG holder.
Finally, a standar Peltier Cell is kept on top of the strip detector in order to stabilize its temperature in the working range of $18^{\circ} - 28^{\circ}$.
The resulting 10-axis motorized positioning system is mounted on a set of Drylin rails and carriers to ensure better stability and alignement and, in addition, to easily adjust source-crystal-detector
positions for each energy to be measured.

\subsection{Target illuminated region}

\begin{figure}[htbp]
\centering
\includegraphics[width=8.5cm]{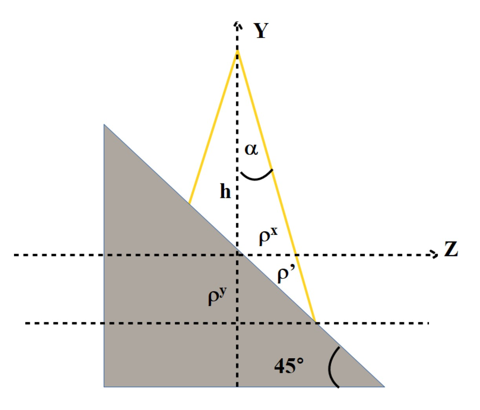}
\caption{Schematics of the illuminated region of the target (not in scale). See text for more details.}
\label{illuminated}
\end{figure}

The region of the target illuminated by the X-ray tube is constant and can be calculated starting from the initial anode position $h = 95,3 \,mm$ 
and the angular conical aperture $\alpha = 11^{\circ}$ of the tube (see Fig. \ref{illuminated}).

\noindent  The intersection of the conical beam and the prism, rotated at
$45^{\circ}$, is resulting to be ellypsoidal with axes:

\begin{equation}
\rho=\rho^x=htg\alpha=95,3 \, mm \times tg(11^{\circ}) = 18,524 \, mm 
\end{equation}

\begin{equation}
\rho'=\frac{\rho sin(\frac{\pi}{2}+\alpha)}{sin(\frac{\pi}{4}-\alpha)} = 32,52 \, mm 
\end{equation}

\begin{equation}
\rho^y=\rho'sin\frac{\pi}{4} = 22,99 \, mm 
\end{equation}

\subsection{Source size in dispersion plane}

\noindent As pointed out in section \ref{intro}, the actual limitations on the possible usage of crystal spectrometers for extended targets is represented by the requirement of a point-like source; however, using a pair of 
slits as the one described in the previous section, it is possible to shape the beam of X-rays emitted by an extended and diffused target in such a way to simulate a virtual point-like source.

\begin{figure}[htbp]
\centering
\includegraphics[width=8.7cm]{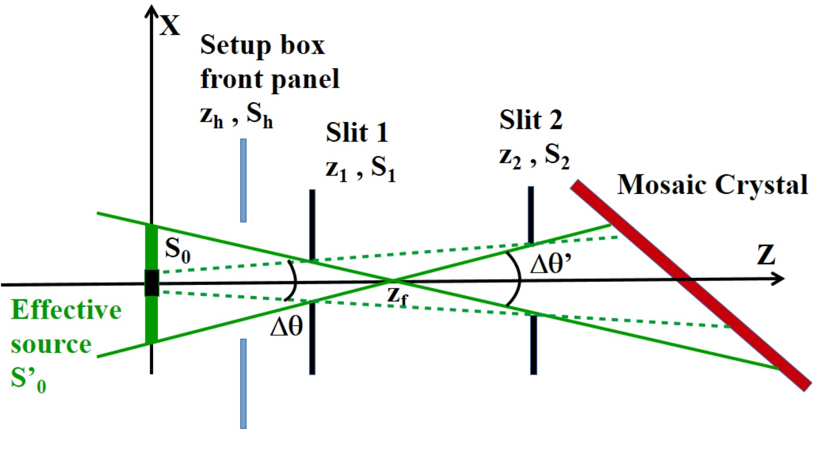}%
\caption{Beam geometry on the dispersive plane (not in scale): the position ($z_1$ and $z_2$) and the aperture ($S_1$ and $S_2$) of two slits are used to create a virtual source between the two slits 
($z_f$, green solid lines), and effective source $S_0'$ (green) and an angular acceptance $\Delta\theta'$; the HAPG crystal is pictured in red, the two slits are shown in black while $z_h$ and $S_h$ are the position and the diameter of the circular exit window of the aluminum box front panel (light blue), respectively.}
\label{horizontal}
\end{figure}

\noindent Referring to Fig. \ref{horizontal}, this configuration is obtained setting the position ($z_1$ and $z_2$) and the aperture ($S_1$ and $S_2$) of each slit in order to create a virtual source between the two slits 
($z_f$, green solid lines on Fig. \ref{horizontal}), an angular acceptance $\Delta\theta'$, and an effective source $S_0'$ (green) which can be, in principle, as wide as necessary. 
The $\Delta\theta'$ angular acceptance could also be set to any value, provided it is large enough to ensure that all the $\theta_B$ corresponding to the lines to be measured are included; for example, if both 
Fe($K_{\alpha1}$) (4510,84 eV, $\theta_B=24,19^{\circ}$) and Fe($K_{\alpha2}$) (4504,86 eV, $\theta_B=24,22^{\circ}$) have to be measured, taking also into account the mosaicity ($\sigma_{\omega}=\omega_{FWHM}/2,35$), the condition is $\Delta\theta'+6\sigma_{\omega}\ge0,03^{\circ}$, since for each photon direction there is a non-zero probability to find a properly oriented crystallite within $6\sigma_{\omega}$.
These X-ray beam shape characteristics overlaps with a second component, depicted as a dotted green line in Fig. \ref{horizontal}, due to the X-rays emitted in the inner part of the target ($S_0$) with a smaller angular
divergence $\Delta\theta \le \Delta\theta'$. In Fig. \ref{horizontal}, $z_h$ and $S_h$ are the position and the diameter of the circular exit window of the aluminum box front panel, respectively.

\begin{figure}[htbp]
\centering
\includegraphics[width=8.7cm]{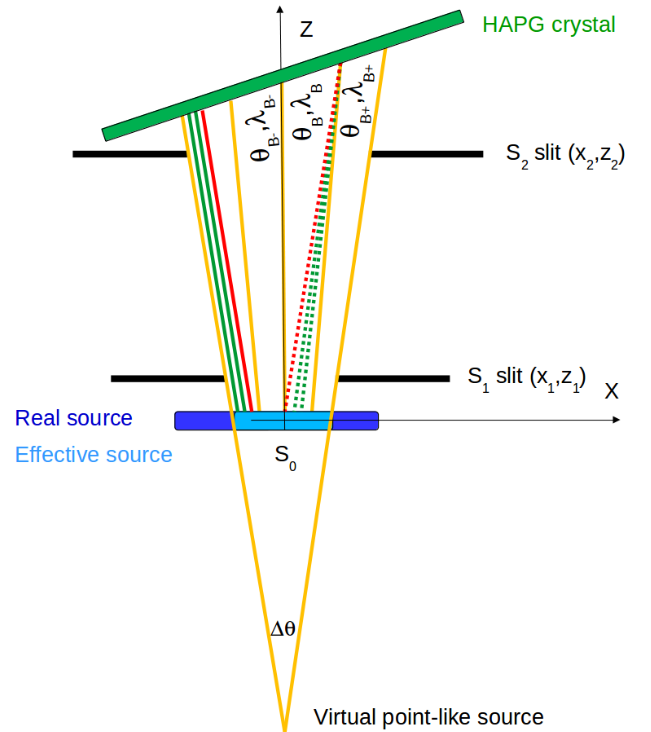}%
\caption{Signal and background scheme (not in scale); two slits S1 and S2 are used to shape the X-ray beam divergence ($\Delta\theta$) and effective source size ($S_0$). 
Depending on the HAPG mosaic spread, photons emitted parallel (solid lines) and not (dashed lines) to the nominal one matching the Bragg condition (yellow) are shown; 
some of them are reflected under the signal peak (solid and dashed green), some are a source of background (solid red), some are not reflected (dashed red). 
See the text for more details. \label{background}}
\end{figure}

\noindent In practice, increasing the effective source size introduces a bakground, as shown in Fig. \ref{background}; for simplicity we describe this situation only for the dotted green lines scheme of Fig. \ref{horizontal}
but it applies also in the solid line case. The yellow lines on the figure represent the photons which, matching the Bragg condition ($\theta_B,\,\lambda_B$), 
form the signal peak on the reflected Bragg spectrum (we call them nominal from now on).
Since the photons are isotropically emitted from the whole target foil, some of them may have the correct energy and angle to be reflected but originate from a point of the 
target near the nominal one (on the yellow line). 
As far as this mislocation is below the limit given by the mosaic spread of the crystal, such photons are also reflected under the 
signal peak worsening the spectral resolution (solid green lines in the figure); 
on the contrary, when this mislocation exceeds this limit (solid red line) these photons are reflected outside the signal peak.
In the same way, photons not emitted in parallel to the nominal ones may still impinge on the HAPG crystals with an angle below its mosaic spread (dashed green lines) and be then 
reflected under the signal peak also affecting the spectral resolution. On the contrary, if the impinging angle is out of this limit, 
those photons are not reflected on the position detector.
As a consequence, for each energy there is the possibility to find the right slits configuration leading at the maximum source size keeping the resolution below the desired limit.

\noindent For each chosen $\Delta\theta',S_0'$ pair, the corresponding values of the slits' apertures $S_1$ and $S_2$ can be found; first, we define the position of the intersection point $z_f$:

\begin{equation}
z_f = \frac{S_0'}{2}ctg\frac{\Delta\theta'}{2}
\end{equation}

\noindent Then, the two slits' apertures and the vertical illuminated region of the HAPG are defined by:

\begin{equation}
S_1 = \frac{z_f-z_1}{z_f}S_0'
\end{equation}

\begin{equation}
S_2 = \frac{z_2-z_f}{z_f}S_0'
\end{equation}

\noindent Concerning the second component, the $\Delta\theta$ and $S_0$ parameters can be otained from $z_1,S_1$ and $z_2,S_2$ as:

\begin{equation}
\Delta\theta = tg^{-1}\frac{S_2-S_1}{z_2-z_1}
\end{equation}

\begin{equation}
S_0 = S_2-2z_2tg\frac{\Delta\theta}{2}
\end{equation}

\subsection{Source size in focusing vertical plane}

\begin{figure}[htb]
\centering
\includegraphics[width=8.5cm]{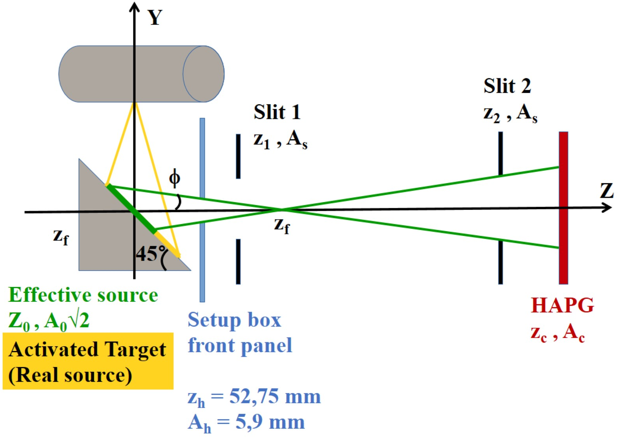}
\caption{Beam geometry on the focusing plane. The vertical spread of the X-ray beam is fixed by the slits positions ($z_1,z_2$) and their frame size ($A_s$), 
together with the exit circular hole in the front panel of the setup box ($z_h,A_h$). See text for more details. \label{vertical}}
\end{figure}

\noindent In Fig. \ref{vertical}, the beam configuration in the vertical plane, orthogonal to the dispersive one, is shown. 
The vertical spread of the X-ray beam is fixed by the slits positions ($z_1,z_2$) and their frame size ($A_s$), 
together with the exit circular hole in the front panel of the setup box ($z_h,A_h$). 
The $\phi$ angle is then defined as:

\begin{equation}
tg\phi = \frac{A_s+A_h}{2(z_2-z_h)}
\end{equation}

\noindent We first define, as in the horizontal case, the position of the intersection point $z_f$:

\begin{equation}
z_f = z_2-\frac{A_s}{2tg\phi}
\end{equation}

\noindent The vertical spread of the beam on the target ($A_0$) and on the HAPG crystal ($A_c$) are then:

\begin{equation}
A_0 =2z_ftg\phi
\end{equation}

\begin{equation}
A_c =2(z_c-z_f)tg\phi
\end{equation}

\noindent where $z_c$ is the crystal position (corresponding to $L_1$ in Fig. \ref{vhl1l2}).

\noindent The final effective source area is then determined by $S_0'\times A_0$.

\section{Experimental Results\label{measure}}

\noindent In this section we present the spectra obtained for different lines (see tab. \ref{angpar}); 
for each measurement, the corresponding geometrical parameters are listed in tab. \ref{param}, where $\theta_B^{set}$ 
is the central Bragg angle value used for the calculation.  
The crystal curvature radius is chosen in order to make a compromise between the energy resolution and the signal rate, since
higher $\rho_c$ leads to longer paths meaning better resolution but higher X-ray absorption from the air. Slits positions $z_1$ and $z_2$ are chosen 
such as to have a vertical dispersion at the HAPG position smaller than the crystal size ($30\, mm$).

\begin{table}
\caption{List of the X-ray lines used in this work and the corresponding Bragg angles $\theta_B$. \label{angpar}}
\begin{ruledtabular}
\begin{tabular}{c|c|c}
Line & E (eV) & $\theta_B$ ($^{\circ}$) \\
Fe($K_{\alpha1}$) & $6403,84$  & $16,77$ \\
Fe($K_{\alpha2}$) & $6390,84$  & $16,81$ \\
Cu($K_{\alpha1}$) & $8047,78$  & $13,28$ \\
Cu($K_{\alpha2}$) & $8027,83$  & $13,31$ \\
Ni($K_{\beta}$)   & $8264,66$  & $12,92$ \\
Zn($K_{\alpha1}$) & $8638,86$  & $12,35$ \\
Zn($K_{\alpha2}$) & $8615,78$  & $12,39$ \\
Mo($K_{\alpha1}$) & $17479,34$ & $6,07$  \\
Mo($K_{\alpha2}$) & $17374,30$ & $6,11$  \\
Nb($K_{\beta}$)   & $18622,50$ & $5,70$  \\
\end{tabular}
\end{ruledtabular}
\end{table}

\begin{table*}
\caption{List of the measurements presented in this work and their main beam parameters. \label{param}}
\begin{ruledtabular}
\begin{tabular}{ c | c | c | c | c | c | c | c }
Line &  $\theta_B^{set}$ ($^{\circ}$) & $\rho_c $ (mm) & $L_1 $ (mm) & $L_2$ (mm) & $z_1$ (mm)  & $z_2$ (mm) & $A_0$ (mm) \\
Fe($K_{\alpha1,2}$) & $16,77$ & $103,4$ & $358,46$ & $343,15$ & $76$   & $257$ &   $14,14$\\	  
Cu($K_{\alpha1,2}$) & $13,28$ & $206,7$ & $900,54$ & $876,33$ & $60$   & $820$ &   $8,09$\\	  
Cu($K_{\alpha1,2}$)+Ni($K_{\beta}$)+Zn($K_{\alpha1,2}$)   &  $12,92$ & $103,4$ & $463,07$ & $451,28$ & $75$   & $352$ &   $11,52$\\	  
Mo($K_{\alpha1,2}$)+Nb($K_{\beta}$) & $6,07$  & $77,5$  & $733,35$ & $729,18$ & $98,7$ & $649,7$ & $13,97$\\	  
\end{tabular}
\end{ruledtabular}
\end{table*}

\subsection{``semi'' - Von Hamos configuration\label{semi}}

\begin{figure}[htbp]
\centering
\includegraphics[width=8.7cm]{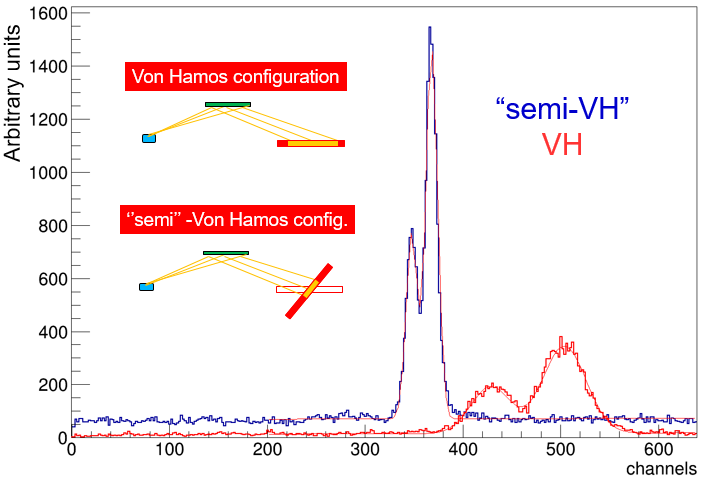}
\caption{Comparison between the standard (red) and the ``semi'' (blue) Von Hamos configuration uncalibrated spectra of Cu($K_{\alpha1,2}$) lines.
In the insight, a schematic of the two configurations is shown where the X-ray source, the HAPG crystal, the position detector and its illuminated region 
are colored in light blue, green, red and yellow, respectively. \label{semi_uncal}}
\end{figure}

\begin{figure}[htbp]
\centering
\includegraphics[width=8.7cm]{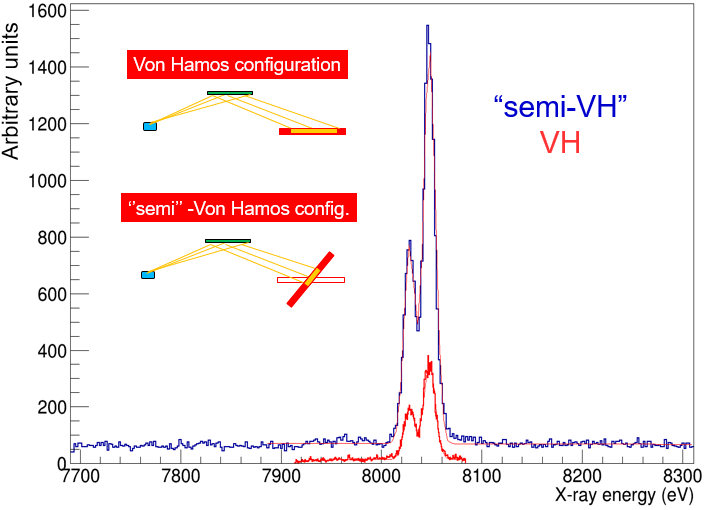}
\caption{Comparison between the standard (red) and the ``semi'' (blue) Von Hamos configuration calibrated spectra of Cu($K_{\alpha1,2}$) lines.
In the insight, a schematic of the two configurations is shown where the X-ray source, the HAPG crystal, the position detector and its illuminated region 
are colored in light blue, green, red and yellow, respectively. \label{semi_cal}}
\end{figure}

\noindent One of the most critical parameter of an X-ray spectrometer is its dynamic range; in particular, the possibility to record multiple lines on a single spectrum is 
crucial if low rates physical processes have to be measured simultaneously and it also allows online calibration.
A possible method to increase the dynamic range is to rotate the position detector of an angle $\theta_M$, like shown in Figs. \ref{semi_uncal} and \ref{semi_cal},
where the comparisons between two measurements in the standard (red) and ``semi'' (blue) Von Hamos configuration are reported.

\noindent The spectra in the main frame are obtained from the MYTHEN2-1D position detector ($50\,\mu m$ per channel), while in the upper left insight the corresponding 
calibrated spectra are shown. From these latter, one can immediately see how the dynamic range (in eV) of the ``semi'' Von Hamos spectrum is wider than the standard one.
In the upper right inside, a schematic of the two configurations is shown where the X-ray source, the HAPG crystal, the position detector and its illuminated region 
are colored in light blue, green, red and yellow, respectively. 

\begin{figure}[htbp]
\centering
\includegraphics[width=8.7cm]{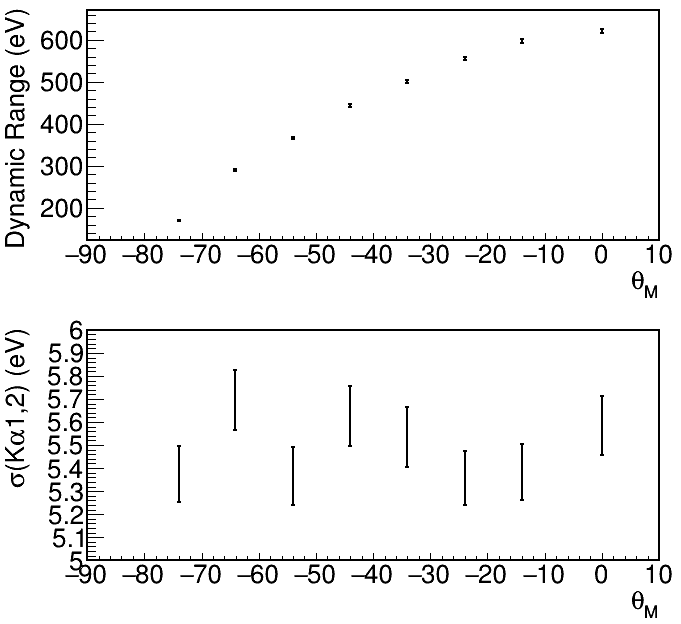}
\caption{Results of the Cu($K_{\alpha1,2}$) lines measured with a $206,7\,mm$ radius HAPG for different rotation angles of the position detector $\theta_M$, where $\theta_M=0$ refers to the ``semi'' Von Hamos configuration; in the top panel the dynamic range (eV), defined as $E_{max}-E_{min}$ of the calibrated spectrum is plotted, while in the bottom one the resolutions ($\sigma$) of the $K_{\alpha1,2}$ peaks are shown. The fitting function is a double gaussian with common $\sigma$ for the Cu lines and a polynomial for the background. In this measurement, $S_0'=1,2\,mm$ and $\delta\theta'=0,5^{\circ}$. }
\label{mythrot}
\end{figure}

\noindent In order to check how the peak resolution is influenced by this rotation, we report as an example in Fig. \ref{mythrot} the results of a set of measurements of the Cu($K_{\alpha1,2}$), for different $\theta_M$ values ranging from $0^{\circ}$ to $-90^{\circ}+\theta_B$, using a $206,7\,mm$ radius HAPG with $S_0'=1,2\,mm$ and $\delta\theta'=0,5^{\circ}$. The $\theta_M=0$ position corresponds to the ``semi'' Von Hamos configuration in which the position detector is rotated in order to have the photons reflected with the nominal Bragg angle $\theta_B$ impinging orthogonally to the detector surface. The fitting function is a double gaussian with common $\sigma$ for the Cu lines and a polynomial for the background. 

\noindent In the top panel the dynamic range, defined as $E_{max}-E_{min}\, (eV)$ of the calibrated spectrum is plotted, while in the bottom one the resolutions ($\sigma$) of the $K_{\alpha1,2}$ peaks are shown. From these results, one can appreciate how the dynamic range is strongly increased while the resolution remains constant within the error bars. This very important result triggered the decision to perform all the subsequent measurements in the ``semi'' Von Hamos configuration.
 
\subsection{Measured spectra}

\begin{figure}[htbp]
\centering
\includegraphics[width=8.7cm]{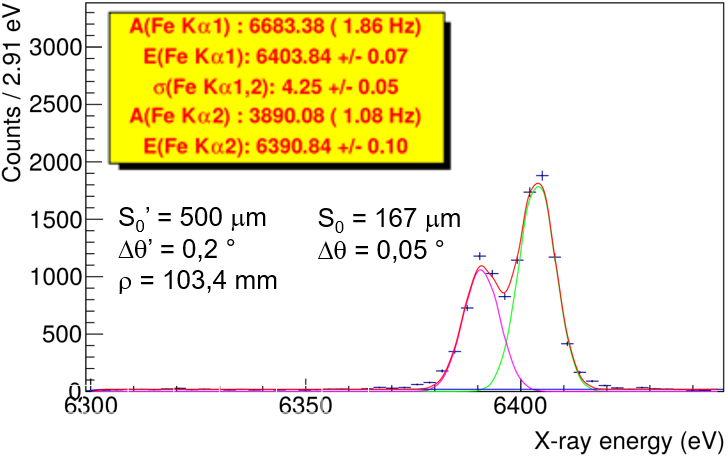}
\caption{Fitted spectrum of Fe($K_{\alpha1,2}$) lines: $K_{\alpha1}$, $K_{\alpha2}$, polynomial background and total fitting function correspond to the green, violet, blue and red curves, respectively.}
\label{fe}
\end{figure}

\begin{figure}[htbp]
\centering
\includegraphics[width=8.7cm]{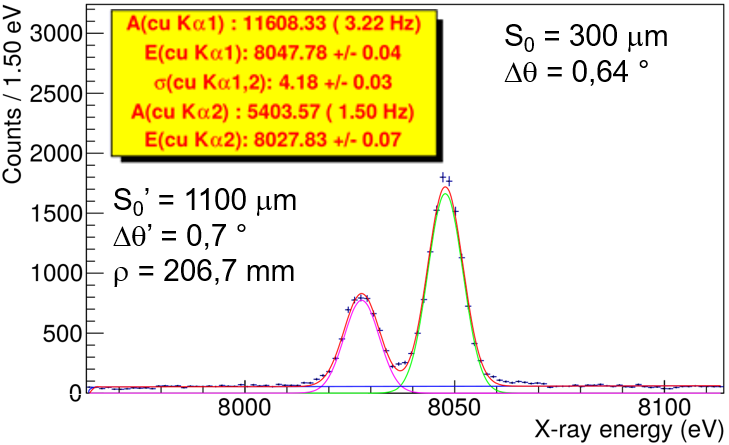}
\caption{Fitted spectrum of Cu($K_{\alpha1,2}$) lines: $K_{\alpha1}$, $K_{\alpha2}$, polynomial background and total fitting function correspond to the green, violet, blue and red curves, respectively.}
\label{cu}
\end{figure}

\begin{figure}[htbp]
\centering
\includegraphics[width=8.7cm]{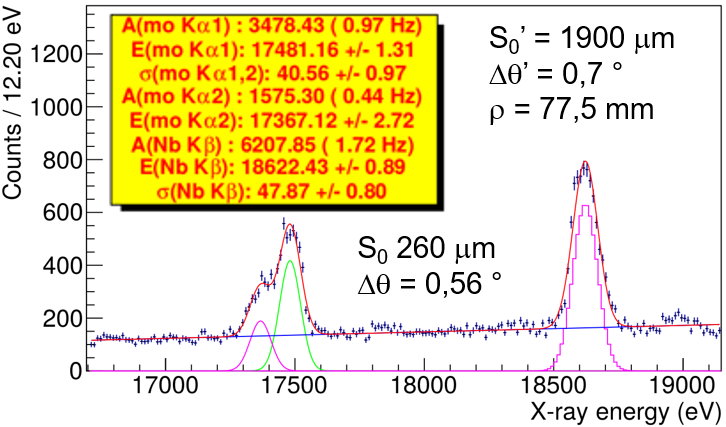}
\caption{Fitted spectrum of Mo($K_{\alpha1,2}$)+Nb($K_{\beta}$) lines: $K_{\alpha1}$, $K_{\alpha2}$ and $K_{\beta}$ fitting function correspond to the green, violet and light blue curves, respectively and are, together with a polynomial background (blue) and the overall fit (red), overimposed to the calibrated spectrum. }
\label{monb}
\end{figure}

\begin{figure}[htbp]
\centering
\includegraphics[width=8.7cm]{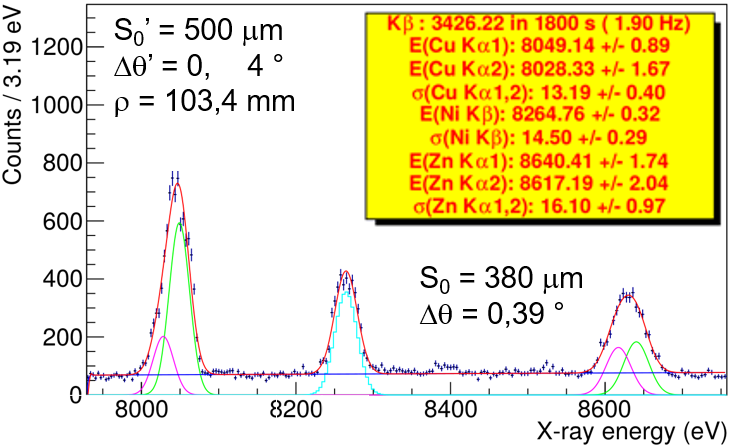}
\caption{Fitted spectrum of Cu($K_{\alpha1,2}$)+Zn($K_{\beta}$)+Ni($K_{\alpha1,2}$) lines: $K_{\alpha1}$, $K_{\alpha2}$ and $K_{\beta}$ fitting function correspond to the green, violet and light blue curves, respectively and are, together with the overall fit (red), overimposed to the calibrated spectrum.}
\label{cuznni}
\end{figure}

\noindent In this section we present the results of the measurements listed in table \ref{param}.
For each measurement, the spectra have been acquired with different $S_0'$ and $\Delta\theta'$ settings but with a fixed integration time; in Figs. \ref{fe}, \ref{cu}, \ref{monb} and \ref{cuznni} we show only one fitted spectrum per measurement corresponding to the best achieved precision on the line position. 
The individual peak fitting functions are gaussians with a common $\sigma$ parameter for $K_{\alpha1,2}$. 
$K_{\alpha1}$, $K_{\alpha2}$ and $K_{\beta}$ fitting functions correspond to the green, violet and light blue curves, respectively and are, together with a polynomial backgound (blue) and the overall fit (red), overimposed to the calibrated spectrum.
The values of the $S_0'$,\,$\Delta\theta'$ pair, the corresponding $S_0$,\,$\Delta\theta$ pair and the curvature radius $\rho_c$ of the used crystal are also reported in the figures.

\noindent The Fe($K_{\alpha1,2}$) measurement has been performed using a smaller curvature radius crystal with respect to the Cu($K_{\alpha1,2}$), since the air
absorption for 6 keV is higher than for 8 keV, so the photon paths (source-crystal and crystal-detector) had to be kept as small as possible. 
In both these measurements the $\Delta\theta'$ angular acceptance could have been set to small values because the two lines of interest were only few percent of degree
distant one from the other; nevertheless, spectra with resolution of $\sigma\simeq\,4\,eV$ have been obtained for both Fe and Cu with effective source sizes of $500\,\mu m$ and $1,2\,mm$, respectively. These values are already one order of magnitude higher than the ones presented in section \ref{intro}. 
The effective source size is even wider in the case of the MoNb measurement at higher energies, where resolutions of $\sigma\simeq40\,eV$ have been obtained using a 
$\rho_c=77,5\,mm$ radius HAPG crystal to avoid the almost $2\,m$ path lengths resulting from bigger $\rho_c$ values. 

\noindent It has to be mentioned that the worsening in resolution could be avoided exploiting the n=2 reflection order; this would correspond to a measurement of a line around $9\,keV$ for n=1, leading to a resolution similar to the one obtained for the Cu target. As a drawback, this would require a longer exposure time since the n=2 peak reflectivity is an order of magnitude smaller than the n=1 one \cite{Gerlach:2015}.
The possibility to exploit a wide dynamic range is highlighted by the CuZnNi spectrum, in which the combined effect of the $\Delta\theta'$ angular acceptance and the mosaicity allow to have in a single measurement lines which are almost $600\,eV$ distant. It has to be noticed that since the lowest and highest part of the spectrum are only exploiting the tail of the mosaicity curve, the Cu($K_{\alpha2}$) and the Zn($K_{\alpha1}$) peaks are a bit suppresed; this is the reason why they don't have the standard 0.5 ratio between $K_{\alpha1}$ and $K_{\alpha2}$.

\section{Conclusions}

\noindent In this paper we presented the results obtained with the VOXES compact Von Hamos spectrometer based on mosaic crystals of pyrolitic graphite (HAPG); in particular, we demonstrated how the proposed spectrometer allows to obtain energy resolutions of few eV when used with extended and diffused isotropic sources. The possibility to use it in a ``semi'' Von Hamos configuration, leading to a wider energy range with no loss in resolution was also confirmed.
Another very important feature of this spectrometer (and of Bragg spectrometers in general) is the possibility to have almost background-free spectra because of the non linear total path of the photons; indeed, putting a soft X-ray shielding of few mm of plastic material along the $L_2$ path will prevent backround scattered photons coming from the activated target to arrive directly on the position detector. A confirmation of this feature can be deduced by the errors obtained in the peak energy positions shown in our spectra, which follow the pure statistical behaviour of a gaussian and background free signal having $\delta E \simeq \frac{\sigma}{\sqrt(N)}$. This, together with a proper shielding around the position detector itself to be adapted to the different hadronic and electromagnetic machine backgrounds, make this spectroscopy technique a serious candidate for high precision X-ray measurements from diffused and extended sources in accelerator environments.

\begin{acknowledgments}
\noindent This work is supported by the 5th National Scientific Committee of INFN in the framework of the Young Researcher Grant 2015, n. 17367/2015.
We thank the LNF and SMI staff, in particular the LNF SPCM service and Doris Pristauz-Telsnigg, for the support in the preparation of the setup.
\end{acknowledgments}

\bibliography{channeling2018-Scordo}

\end{document}